\documentclass[prb,twocolumn,showpacs,preprintnumbers,amsmath,amssymb]{revtex4}
\usepackage{graphicx}% Include figure files
\usepackage{dcolumn}% Align table columns on decimal point
\usepackage{bm}% bold math
\usepackage{psfig}
\newcommand \be{\begin{eqnarray}}
\newcommand \ee{\end{eqnarray}}
\newcommand{\ket}[1]{\left|#1\right\rangle}
\newcommand{\bra}[1]{\left\langle#1\right|}
\begin{document}
\title{
%Charge ratchets: 
Transport and noise in organic field effect devices} 
\author{K. Morawetz$^{1,2,3}$, S. Gemming$^{1}$, R. Luschtinetz$^{4}$, T. Kunze$^{1}$, P.~Lipavsk\'y$^{5,6}$, L. M. Eng$^{7}$, G. Seifert$^{4}$,  V. Pankoke $^{1}$, P. Milde$^7$ 
}
\affiliation{$^1$Forschungszentrum Dresden-Rossendorf, PF 51 01 19, 
01314 Dresden, 
Germany}
\affiliation{$^2$ International Center for Condensed Matter Physics, 
%Universidade de Bras\'ilia, 
70904-910, Bras\'ilia-DF, Brazil}
\affiliation{$^3$Max-Planck-Institute for the Physics of Complex
Systems, N\"othnitzer Str. 38, 01187 Dresden, Germany}
\affiliation{$^4$Institute of Physical Chemistry and Electrochemistry, TU Dresden, 01062 Dresden, Germany}
\affiliation{
$^5$Faculty of Mathematics and Physics, Charles University, Ke Karlovu 3, 12116 Prague 2, Czech Republic}
\affiliation{
$^6$Institute of Physics, Academy of Sciences, Cukrovarnick\'a 10, 16253 Prague 6, Czech Republic
}
\affiliation{$^7$Institute of Applied Photophysics, TU Dresden, 01062 Dresden, Germany}
\begin{abstract}
The transport and fluctuation properties of organic molecules ordered parallel between two Au contact leads are investigated by the method of surface Green function. From first-principles simulation the relevant hopping parameters are extracted and used to calculate nonlinear transport coefficients with respect to an external bias voltage. A staggering of conductance is found in dependence on the number of molecules squeezed in-between the contacts.
The thermal properties show an anomalous behavior whenever the voltage reaches the values of the molecular energy levels active for transport. The thermoelectric figure of merit shows a resonance allowing to reach values even larger than one.
\end{abstract}
\pacs{73.63.Fg, %	Nanotubes
73.23.-b,% 	Electronic transport in mesoscopic systems
85.85.+j,% 	Micro- and nano-electromechanical systems (MEMS/NEMS) and devices
87.15.hj,% 	Transport dynamics
05.60.Cd% 	Classical transport
}
\maketitle

\section{Introduction}

The goal to develop low-cost storage and microelectronic devices has triggered an enormous activity in the research of organic field effect transistors (OFETs) based on different polymers \cite{YT05,KMHSTWF06,HJAKLEGHS08} and small organic molecules \cite{ss08}. It is desirable that the molecular material possesses a high structural ordering \cite{DM02,Ga99} in order to reach high charge carrier mobilities and low resistive losses. Among the most promising materials are oligo\-thiophenes and their derivatives \cite{Fa07}. This is  due to the variety of intra- and intermolecular interactions, which originates from the polarizability of the sulfur electrons and the embedding aromatic $\pi$-electron system \cite{MGBCBMO03,MGTB00,BZBA93}. 

\begin{figure}[h]
\psfig{file=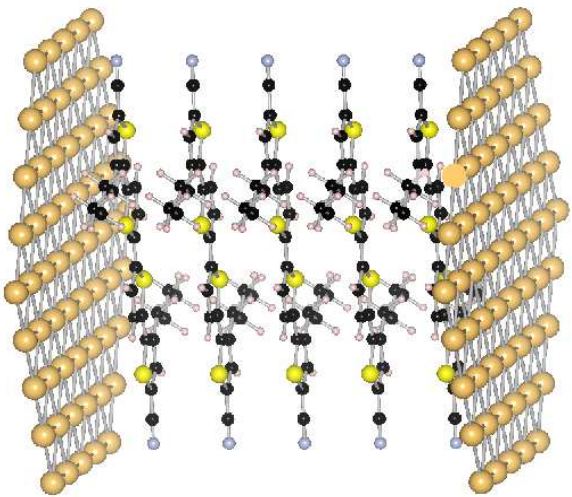,width=8cm}
\caption{The thiophene molecules between 2 Au contacts.\label{bild1}}
\end{figure}

\begin{figure}
\psfig{file=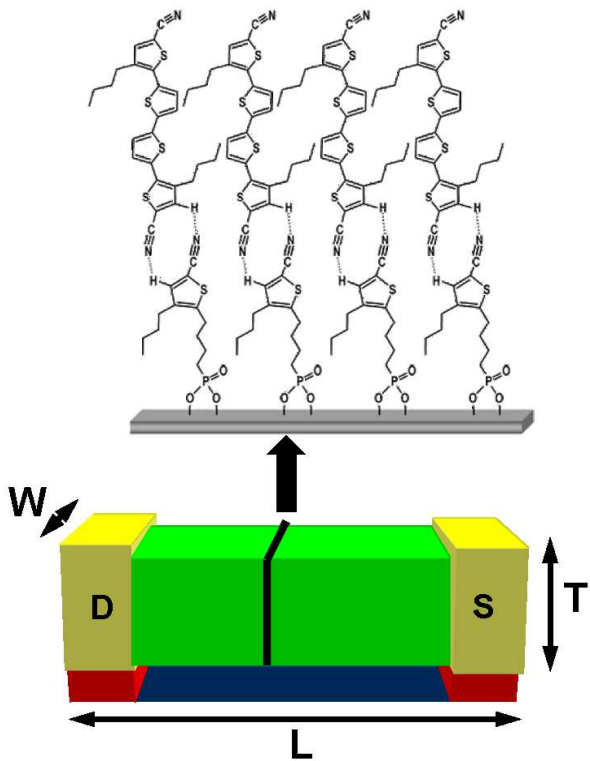,width=8cm}
\caption{The parallel stacking of thiophene molecules between 2 Au contacts building the source (S) and drain (D) below. Typical sizes are $W=1000\mu$m, $T=20$nm and $L=10\mu$m. One slice is enlarged (above) presenting the chemical structure  \cite{HJAKLEGHS08}.
\label{bild2}}
\end{figure}

The performance of OFETs based on oligothiophenes \cite{HJAKLEGHS08} shows characteristic features, e.g. the current starts at a certain threshold of gate voltage and reaches a saturation value for certain drain voltages. 
In general, the charge transport occurs in the direction perpendicular to the plane of the thiophene rings due to good $\pi$-stacking. This suggests to construct devices with parallelly ordered molecular rings. This is the perpendicular direction to the usually considered transport through various molecule classes ranging from transport dependent on the thickness \cite{CGWS05} to metallic behaviour independent of the thickness of molecules\cite{CB07}. Thick-film devices based on thiophenes were reported several years ago  \cite{YDJIC02}. Recently an OFET structure has been built from ultra-thin self-assembled films made up from 
%$\alpha$, $\omega$- and $\beta$, $\beta'$-substituted 
oligothiophenes, which are arranged in a highly-order lamellar stacking perpendicular to the substrate surface \cite{HJAKLEGHS08} as illustrated in figures~\ref{bild1} and \ref{bild2}. 

The energy gap between the lowest unoccupied (LUMO) and the highest occupied molecular level (HOMO) measures about $3$ eV in quarterthiophene. For the molecular structure see figure~\ref{bild1}. 
In contrast to the conduction and valence band in semiconductors the transport is due to localized states dominated by hopping \cite{H98,F07} rather than due to delocalized states limited by the scattering in semiconductors. 

If one considers the charge transport between the molecules one can think of two extreme mechanisms. On one side one has either the ballistic transport through bonds, or the hopping transport through space. The diffusive transport on the other side is present if the carriers meet scattering partners. For the present situation of parallely stacked molecules the hopping transport by tunneling between the localized HOMO or LUMO states is the relevant one. A related mechanism of charge transfer due to the bending motion of molecules by touching might lead to shuttling transport \cite{MGLESK08} which could be investigated by resonant spectroscopy \cite{RC07}. Of course, the most crucial question is the tilting of molecules triggered by the metal-molecul contacts \cite{GSWC06} which requires a refined interface engineering \cite{PLLC07}.

Here we construct a hopping transport picture and want to investigate the particle and thermal transport properties as well as current fluctuations. We will project the electron transport including Coulomb and inelastic tunneling to a simple one-particle hopping Hamiltonian by calculating the matrix elements through overlap integrals of orbitals obtained from density functional calculation. We account for the Coulomb force only in determining these effective hopping elements. The Coulomb interaction is not considered explicitly in the hopping Hamiltonian such that we concentrate only on coherent motion but under nonlinear external bias voltage. The Coulomb effect to the hopping transport has been discussed e.g. in [\onlinecite{KHK06,SRC07,S08}]. One might expect that the Coulomb effect will lead to a blocking of the transport due to the creation of a Coulomb gap. This is beyond the scope of the paper, and since we deal with a low charge per molecule of $\sim 10^{-6}$, we neglect this effect at room temperature. Other inelastic tunneling processes we do not account explicitly in the hopping
model than in the cumulative sense of DFT based hopping matrix elements. Therefore this model is an effective one and cannot claim to resolve detailed information on the single molecular level. The predictive power of such an effective model can finally be checked only with corresponding experimental data which are expected in nearest future.

In calculating the transport properties for hopping situation the Greenfunctions are a convenient tool \cite{CCNS71,CCLNS71,WJM93,JWM94,BB00,HCYMS08}. We will use the method of surface Greenfunctions \cite{SH91,SHLS91}.
Alternative methods of direct numerical inversions or expansions with the help of the Floquet theorem  can be found in the literature, e.g. \cite{KLH05}.
The method of surface Greenfunctions has the advantage to be numerically fast (linear proportional to site numbers) and free of large compensations. Furthermore it allows to describe a time-dependent response in a convenient way \cite{SH91} and can be easily extended to multiband problems \cite{SoL94}.

First we give the basic ideas of the surface Greenfunction method in chapter II with the approach to the current and current fluctuations. The technical details of the numerically extremely fast method are presented in the appendix. In chapter III we describe the first-principles approach to derive the necessary hopping parameters. The numerical results are presented in chapter IV including the nonlinear transport coefficients leading to an anomaly in the thermal transport. The summary discusses possible experimental verifications and the status of current theoretical developments.

\section{Method of surface Greenfunction}
\subsection{Convenient formulation of the boundary problem} 
We consider the tight-binding Hamiltonian
\be
H=\sum\limits_{j=-\infty}^{\infty} \biggl ( \ket{j}v_j\bra{j}+\ket{j\!+\!1}t_{j+1}\bra{j}+\ket{j\!-\!1}t_{j}\bra{j} \biggr )
\label{ham}
\ee
describing $N$ molecules in-between the crystal leads at the left, $j\le 0$, and the right side, $j\ge N+1$. The energy levels are $v_j$ and the hopping between neighbouring sites are $t_j$.

For the formulation of the scattering one might be tempted to use the Lippmann-Schwinger formulation. To this end one writes the
Hamiltonian as a sum of the unperturbed crystal, $H^0$, made by an
extension of the left semiinfinte crystal all over the whole space, and
the perturbation, $H^\prime=H-H^0$. An incoming wave $\psi^0$ is not yet influenced by the barrier of the leads, thus it is the eigenstate of $H^0$,
\begin{equation}
H^0\psi^0=E\psi^0.
\label{ap0d}
\end{equation}
Eigenstates of the homogeneous system are plane waves with a momentum $k$ related to the energy
by $E=2 v_0-2 v_0\cos(k a)$ with the distance of atoms $a$. The wave approaching the barrier is $\psi^0(x)=
\exp(ikx)$ or discretized $x_j=j\,a$ in the tight-binding representation $\psi^0_j=\exp(ikaj)$.

The total wave function $\psi^0+\psi'$ solves Schr\"odinger's equation
\begin{equation}
(H^0+H^\prime)(\psi^0+\psi')=E(\psi^0+\psi'),
\label{ap1}
\end{equation}
which can be rearranged with the help of (\ref{ap0d}) into the
Lippmann-Schwinger equation,
\begin{equation}
(E-H)\psi'=H^\prime\psi^0.
\label{ap3}
\end{equation}
The boundary condition that
the diffracted wave propagates from the barrier to infinity is realized
by an infinitesimal shift of the energy $E$ into the complex plane.

The above straightforward implementation of the Lippmann-Schwinger idea
becomes numerically extremely inconvenient if the left and right leads are 
different by hopping, energy levels and external bias. For
instance, when the tunneling junction is biased by the voltage $V$, the
perturbation $H^\prime$ in the right half-space equals the potential
of the bias voltage, $H^\prime_{jj}=v_{N+1}$ for all $j\geq N+1$. The
perturbation then extends everywhere except for the left lead, and the
source term in the right hand side of (\ref{ap3}) is nonzero over this
region, too.

In this case, the Lippmann-Schwinger separation of the wave function
into an infinite plane wave $\psi^0$ and correction $\psi$ is not a
favourable starting step. This is clearly seen for the case when electrons
have an energy at the level of the right lead. All electrons are then
reflected at the barrier and only exponential tails of the wave function
penetrate into the lead. The finite value of $\psi^0$ has to be
compensated by $\psi$. This requires a large source term,
$H^\prime\psi^0$, and a very accurate treatment of the
Lippmann-Schwinger equation. In fact, only identities can guarantee a
correct compensation of the large parts of incoming and outgoing waves.

There is a simple modification of the Lippmann-Schwinger idea by which we
can circumvent the penetration of the incoming wave into the barrier
and the right lead. Let us cut the incoming wave by putting
\begin{eqnarray}
\psi^C_j=&\psi^0_j&\ \ \ {\rm for} \ \ j\leq 0,\ \ {\rm and}
\nonumber\\
\psi^C_j=&0&\ \ \ {\rm for} \ \  j\geq 1.
\label{ap4}
\end{eqnarray}
The wave function is now split differently than before, $\psi^0+\psi'=\psi^C+\psi$, and Schr\"odinger's
equation, $(E-H)(\psi^C+\psi)=0$, gives
\begin{equation}
(E-H)\psi=\Omega, \ \ \ \ \ \ \ \ \Omega=-(E-H)\psi^C.
\label{ap5}
\end{equation}
The term $\Omega$ is a well behaved source which is nonzero only at two
layers, $j=0$ and $j=1$, with
\begin{equation}
\Omega_0=-t_1\psi^0_1=-t_1{\rm e}^{ik a}, \ \ \ \ \ \ \ \ \Omega_1=t_1\psi^0_0=
t_1, 
\label{ap8}
\end{equation}
which can be easily proven calculating $\bra{j} E-H\ket{\psi^C}$ with the help of (\ref{ham}). The wave function $\psi$ does not include the incoming part, we can thus
use the retarded boundary condition, $z=E+i0$, and write the wave
function in terms of Greenfunction,
\begin{equation}
\psi={1\over z-H}\Omega=G\Omega
\label{ap11}
\end{equation}
or explicitly
\be
\psi_j=G_{j1}\Omega_1+G_{j0}\Omega_0.
\ee

\subsection{Reflection and transmission}

When considering the left lead, the incoming wave shows the reflected part characterized by the reflection coefficient $r$
\be
\psi_{-1}&=&r{\rm e}^{i k a}
\nonumber\\
&=&t_0 S_{-1}^l (G_{01}\Omega_1+G_{00}\Omega_0)
\label{psl}
\ee
where we used the recursion formula (\ref{gb}) for the Greenfunction. Since the energy conservation reads $\cos{(ka)}=(z-v_0)/2 t_0$ we employ the explicit from of the surface Greenfunction in the lead (\ref{s0}) to derive the relation 
\be
t_0 S_0^l={\rm e}^{ika}.
\ee
This allows us to identify from (\ref{psl}) the reflection coefficient
\be
r=G_{01}\Omega_1+G_{00} \Omega_0.
\ee
Considering the right lead we have the transmission coefficient according to
\be
\psi_{N+1}&=&t {\rm e}^{i\kappa a (N+1)}
\nonumber\\
&=&G_{N+1,1}\Omega_1+G_{N+1,0}\Omega_0
\nonumber\\
&=&t_{N+1} S_{N+1}^r (G_{N,1}\Omega_1+G_{N,0}\Omega_0)
\label{psr}
\ee
where we used the recursion relation (\ref{ga}).
Since the energy conservation on the right lead means $\cos{(\kappa)} a=(z-v_{N+1})/2 t_{N+1}$ we employ the explicit form of the surface Greenfunction in the lead (\ref{sn}) to obtain the relation 
\be
t_{N+1} S_{N+1}^r={\rm e}^{i\kappa a}
\ee
and from (\ref{psr}) follows the transmission coefficient
\be
t&=& {\rm e}^{-i \kappa a N} (G_{N,1}\Omega_1+G_{N0} \Omega_0)
\nonumber\\
&=&-2 i t_1 {\rm e}^{-i \kappa a N} G_{N,0}\, \sin{( k a)}
\nonumber\\
&=&-2 i t_1 {\rm e}^{-i \kappa a (N+1)} G_{N+1,0}\, \sin{( k a)}
\ee
representing different equivalent forms. The reflection $R=|r|^2$ and transmission $T=|t|^2$ obey $T+R=1$, of course.

\subsection{Current and current fluctuations}

For the tight-binding Hamiltonian (\ref{ham}) the velocity operator $\dot x={i\over \hbar} [H,x]$ is easily computed from the position operator $x=a \sum\limits_j j \ket{j}\bra{j}$ as
\be
\dot x ={i a\over \hbar} \sum\limits_j t_{j+1} \left ( \ket {j}\bra{j+1}-\ket {j+1}\bra{j}\right ).
\ee
The mean current $j=\langle e \dot x \rangle$ is conserved. Therefore it is sufficient to calculate the current which runs from the molecules to the lead on the right side
\be
j&=&{i e a \over \hbar} t_{N+1} \left ( \langle \psi^*_N\psi_{N+1}\rangle- \langle \psi^*_{N+1}\psi_{N}\rangle \right )
\nonumber\\
&=& {i e a \over \hbar} t_{N+1} \left ( G^<_{N,N+1}-G^<_{N+1,N}\right )
\label{j}
\ee
where we introduced the definition of the correlated Greenfunctions $G^<$. The Greenfunction provided in the appendix are causal functions. According to the Langreth/Wilkins rules \cite{LW72} which establish the generalized Kadanoff and Baym formalism \cite{LSV86,LSM97} from the product of causal functions $A$ and $B$ one obtains the correlated part via the rule $(AB)^<=A^RB^<+A^<B^A$. The retarded and advanced functions obey the relation $i (A^R-A^A)=(A^<-A^<)$ for fermions. With the help of these rules and the recursion relation of the causal Greenfunction (\ref{ga}) we can write  
\be
G_{N,N+1}^<&=&t_{N+1} (G_{NN}^RS_{N+1}^<+G_{NN}^< S_{N+1}^A)
\nonumber\\
G_{N+1,N}^<&=&t_{N+1} (S_{N+1}^RG_{NN}^<+S_{N+1}^<G_{NN}^A)
\ee
and repeatedly applying (\ref{sl}) the current (\ref{j}) reads
\be
j={a e\over \hbar} \prod\limits_{j=1}^{N+1} t_j^2 \prod\limits_{k=1}^{N-1} |S_k^l|^2 \left (S_{N+1}^{r<}S_0^{l>}-S_{N+1}^{r>}S_0^{l<} \right ).
\label{j1}
\ee
Further simplification can be achieved by rewriting the diagonal Greenfunction $G_{NN}$ in (\ref{j1}) into the Greenfunction $G_{N+1,0}$ with the help of (\ref{gb}). We observe that the surface correlated Greenfunction $S_0^{l<}$ is the one of the left lead. It can be expressed in terms of the spectral function $A_l=i(S_0^{lR}-S_0^{lA})$ and the Fermi-Dirac function $f_l$ describing the occupation of the left lead via $S_0^{l<}=A^l f_l$ and $S_0^{l>}=A^l (1-f_l)$ and analogously for the right lead. One obtains finally for the current $J=j/a$
\be
J(z)&=&{e \over \hbar } |G_{N+1,0}(z)|^2 t_0^2 t_{N+1}^2 A_{N+1}^r(z) A_0^l(z)\nonumber\\
&&\times \left [f_r \left ({z-eV\over T} \right )-f_l\left ({z\over T}\right )\right ].
\label{jf}
\ee 
In order to obtain the conductance we have to integrate over all energies and to divide by the voltage $V=U/e$
\be
G&=&{e\over U} \int \!\!{dz\over 2 \pi} J(z) \equiv {e^2\over U} \int \!\!{d z\over 2 \pi\hbar } T_{N+1,0} \left [ f_r-f_l\right ]
\label{G}
\ee   
which is the known Landauer-B\"uttiker form of conductance \cite{BB00,KLH05}.

The time-averaged current autocorrelation function \cite{CKH04}
\be
{\bar J^2}=\lim\limits_{T\to \infty} {1\over T} \int\limits_0^T dt \int\limits_{-\infty}^\infty d\tau \langle \Delta J(t)\Delta J(t-\tau)\rangle
\ee
describes the zero-frequency noise of the current (\ref{jf}). The result reads
\cite{Bu92,BB00}
\be
{\bar J^2}&=&{e^2\over 2 \pi\hbar U} \int d z T_{N+1,0} \biggl [ T_{N+1,0}(f_r-f_l)^2\nonumber\\&& \qquad \quad +(1-f_r)f_r+(1-f_l)f_l \biggr ].
\ee
As a measure for the fluctuations we will use the Fano factor
\be
F={{\bar J^2}\over e J}.
\label{fano1}
\ee

\section{Calculation of the tight-binding parameters}

Projecting the transport problem to a nearest neigbour hopping we do need to calculate appropriate hopping parameter $t_j=T_{j,j+1}$.
The electronic coupling matrix elements $T_{ij}$ are computed 
using a fragment orbital approach within the self-consistent 
charge density-functional based tight-binding (SCC-DFTB) method 
\cite{Elstner1998,Porezag1995,Seifert1996,Elstner2006,
Seifert2007,Elstner2007}. A similar approach has been already 
successfully applied to calculate charge-transfer matrix 
elements for hole transfer in DNA \cite{Kubar2008_1,Kubar2008_2}.

The electronic coupling between two molecular orbitals $\phi_i$ 
and $\phi_j$ of neighboring molecules or molecular fragments 
$i$ and $j$ is calculated from isolated molecules by the hopping matrix elements
\begin{equation}
T_{ij}=\left\langle \phi_i | \hat H_{KS} | \phi_j \right\rangle
\label{Tij}
\end{equation}
where $H_{KS}$ is the Kohn-Sham Hamiltonian \cite{Kohn1965}.
By applying the atomic-basis-set expansion
\begin{equation}
\phi_i = \sum_{\mu} c_\mu^i \eta_\mu, 
\label{lcao}
\end{equation}
the coupling integrals in the molecular-orbital basis can be 
efficiently evaluated as 
\begin{equation}
T_{ij}= \sum_{\mu\nu} c_\mu^i c_\nu^j \left\langle \eta_\mu 
| \hat H_{KS} | \eta_\nu \right\rangle = 
\sum_{\mu\nu} c_\mu^i c_\nu^j H_{\mu\nu}.
\label{tij_lcao}
\end{equation}
$H_{\mu\nu}$ is the Hamilton matrix represented in the atomic 
basis. This matrix and the atomic coefficients $c_\mu^i$ are 
estimated using the SCC-DFTB method \cite{Elstner1998,
Porezag1995,Seifert1996,Elstner2006,Seifert2007,Elstner2007}.

\begin{figure}
\psfig{file=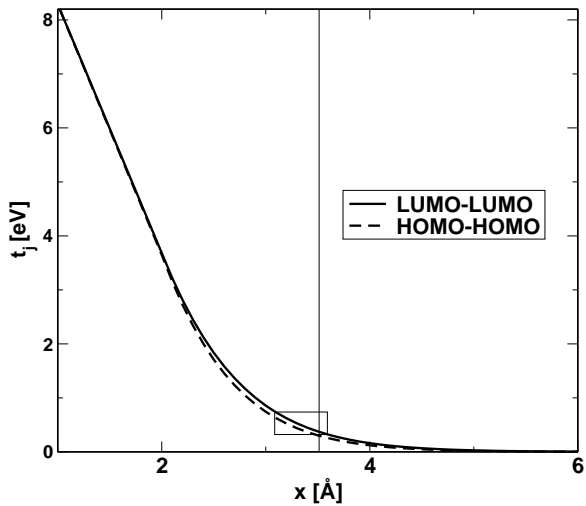,width=9cm}
\caption{The tight binding parameter for nearest neighbor hopping of thiophene versus the distance. The equilibrium distance is marked by a vertical line. The variance of the distance of molecules due to thermal motion is indicated as a box.\label{tij}}
\end{figure}

The SCC-DFTB method is based on the density-functional theory 
of Hohenberg and Kohn \cite{Hohenberg1964} in the formulation 
of Kohn and Sham \cite{Kohn1965}. The single-particle Kohn-Sham
eigenfunctions $\phi_i$ are expanded in a set of localized 
atom-centered basis functions $\eta_\mu$ (Eq.\ref{lcao}). These 
functions are determined by self-consistent density-functional 
calculations on the isolated atoms employing a large set of 
Slater-type basis functions which obey
\begin{equation}
\left[ -{1\over 2} \nabla^2 +v_{eff}[\rho_\alpha]\right] \eta_\mu 
= \epsilon_\mu \eta_\mu. 
\label{dft_ks}
\end{equation}
The calculation of the $T_{ij}$ according to (\ref{Tij})
is based on orthogonal basis functions, 
\begin{equation}
S_{\mu\nu}=\left\langle \tilde{\eta}_\mu |\tilde{\eta}_\nu 
\right\rangle = \delta_{\mu\,\nu},
\label{overlap}
\end{equation}
where the atomic basis functions  at different atomic centers have 
been orthogonalized using the Schmidt orthogonalization
\begin{equation}
|\tilde{\eta}_\mu \rangle = |\eta_\mu \rangle -{1\over2} 
\sum_\nu|\eta_\nu \rangle \langle \eta_\nu |\eta_\mu \rangle.
\end{equation}
Hence, the Hamilton matrix 
\begin{equation}
H_{\mu\nu} =  \left\langle \tilde{\eta_\mu}|\hat{H}_{KS-DFTB}|
\tilde{\eta}_\nu \right\rangle
\label{ham1}
\end{equation}
can be calculated for all necessary combinations of orbitals 
$\tilde{\eta}_\mu$ and $\tilde{\eta}_\nu$ on atoms $\alpha$ and 
$\beta$, and stored in tables.

First, the isolated molecule $i$ has been geometry-optimized 
and the atomic coefficients $c_\mu^i$ of the respective orbital 
$\mu$ have been estimated by standard DFTB calculations 
\cite{Elstner1998,Porezag1995,Seifert1996,Elstner2006,Seifert2007,
Elstner2007}. Second, the atomic Hamilton matrix (\ref{ham1}) 
is constructed in order to calculate the $T_{ij}$ using the 
orthogonalized and non-confined atomic basis functions 
$\tilde{\eta}_\mu$. 

The $T_{ij}$ have been computed between the highest occupied
orbitals (HOMOs) and lowest unoccupied orbitals (LUMOs) of 
two quarterthiophene molecules. The respective nearest neighbour hopping
parameters $t_j=T_{j,j+1}$ in 
dependence on the stacking distance of the molecules is shown 
in figure~\ref{tij}.

\begin{figure}
\psfig{file=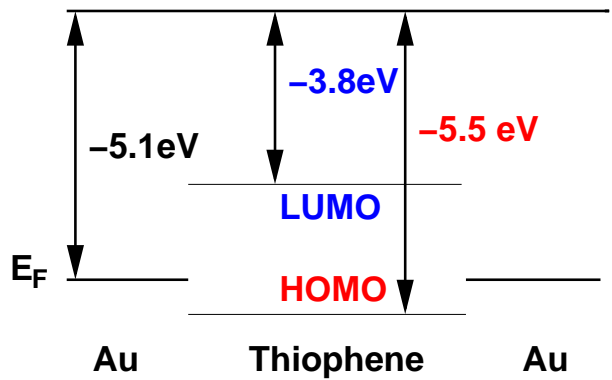,width=8cm}
\caption{The work  function of thiophene molecules with respect to Au contacts. \label{thio}}
\end{figure}

\section{Results}

\subsection{Conductance for thiophene molecules}

\begin{figure}
\psfig{file=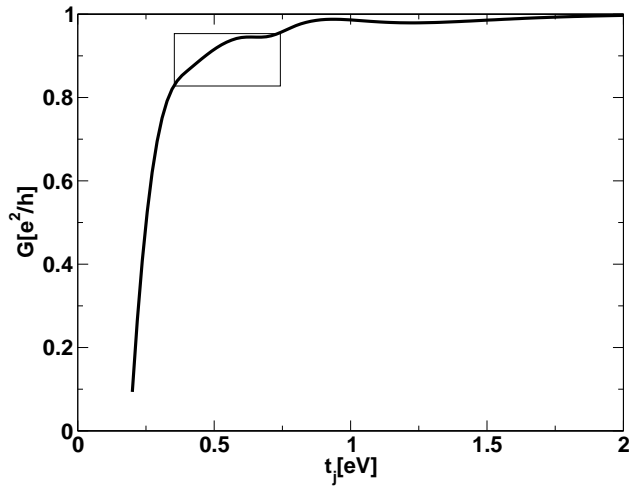,width=9cm}
\caption{The conductance of 20 thiophene molecules between Au contacts versus the tight binding parameter. The variance according to figure \protect\ref{tij} is indicated by the box. \label{homo_t}}
\end{figure}

We present now the results for thiophene molecules between Au contacts. The work function of Au is $-5.1$ eV and the HOMO levels of thiophene are at $-5.5$ eV such that $v_j=-0.4$ eV as illustrated in figure~\ref{thio}. The LUMO levels are much higher at $-3.8$ eV and do not play any role for the conductance here. The calculated hopping parameters of the last section vary slightly with the distance of the molecules as indicated by the  box in figure~\ref{tij}. This translates into a slight variation of the conductance as well. We have plotted in figure~\ref{homo_t} the dependence of the linear conductance at infinitesimal bias on the hopping parameter. A variance of the distance of the molecules between 2 and 4 \AA{} results only in a variation of the conductance by around $10\%$ due to the flat curves in figures \ref{tij} and \ref{homo_t}.

\begin{figure}
\psfig{file=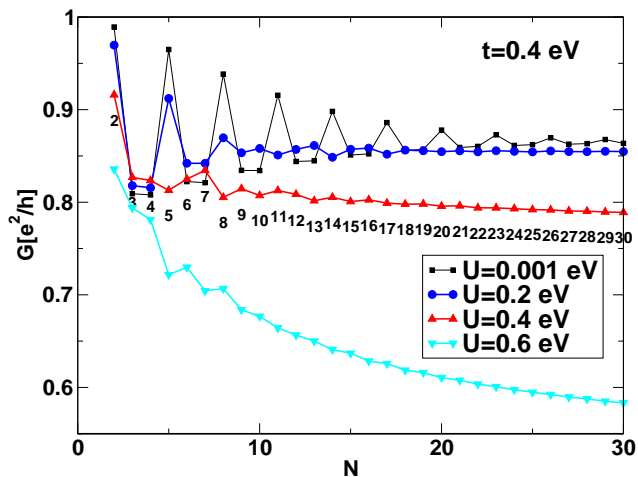,width=9cm}
\caption{The conductance versus the number of thiophene molecules between Au contacts for different applied voltages.\label{homo_N}}
\end{figure}

It is interesting to discuss the dependence of the conductance on the number of molecules squeezed in-between the contacts. In figure~\ref{homo_N} we plot the conductance as a function of the number of molecules for different applied voltages. One recognizes a staggering up to 20-25 molecules. Above this critical number the conductance saturates. With increasing applied voltage this staggering is already damped out for a smaller number of particles besides the overall decreasing of the conductance which will be discussed below.

\begin{figure}
\psfig{file=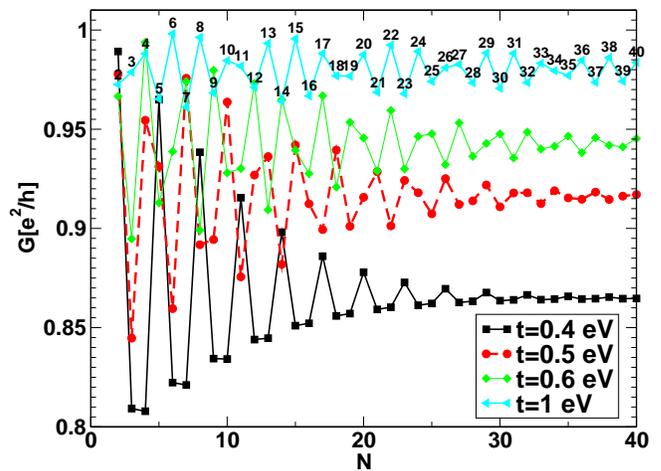,width=9cm}
\caption{The conductance versus the number of thiophene molecules between Au contacts for different tight binding parameter in the linear response regime of applied voltage.\label{homo_N-t3}}
\end{figure}

In order to find the origin of this effect let us plot the conductance for different tight binding parameter. In figure~\ref{homo_N-t3} we see that the overall conductance becomes larger for larger tight binding parameter which is obvious due to the better coupling between the molecules. Moreover one can see that the period of staggering becomes shorter for larger tight binding parameter. This suggests that we possibly see here a coherence effect over a certain number of molecules which is controlled by the hopping parameter.

\subsection{Density of states, current and current fluctuation}

The conductance is sensitively dependent on the relative hopping parameter in the Au lead compared to the molecule ones. In the next figure~\ref{g_t0} the hopping parameter in the Au leads are varied and the nonlinear conductance is plotted versus applied voltage bias. 

\begin{figure}
\psfig{file=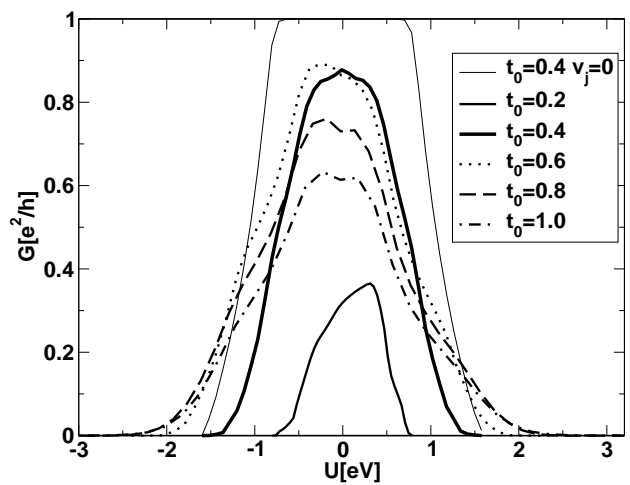,width=9cm}
\caption{The conductance versus applied voltage bias for different hopping parameters from the lead to the thiophene. The energy levels of the leads are at $v_0=0$ and the HOMO energy levels of the thiophene are at $v_j=-0.4$. As reference curve the case of homogeneous lead material, $v_j=0$, is plotted as well (thin line).\label{g_t0}}
\end{figure}

Compared with the homogeneous case of the lead material, $v_0=v_j=0$, which gives the maximal conductance, one sees that the mismatch between the lead hopping parameter and the molecule hopping parameter lowers the conductance. While a smaller hopping parameter in the leads lead to an overall shrinkage of the current-voltage curve, a large hopping parameter in the lead leads to a lower but broader curve. This means that the device is conductive even for higher applied voltages where the homogeneous case does not allow a current any more since the energy dispersions of the left and right layer do not overlap. This is of course a mechanism which can be used as molecular switch \cite{VGTC07,PGORS08}.

\begin{figure}
\parbox[]{9.5cm}{
\parbox[]{4.6cm}{
\psfig{file=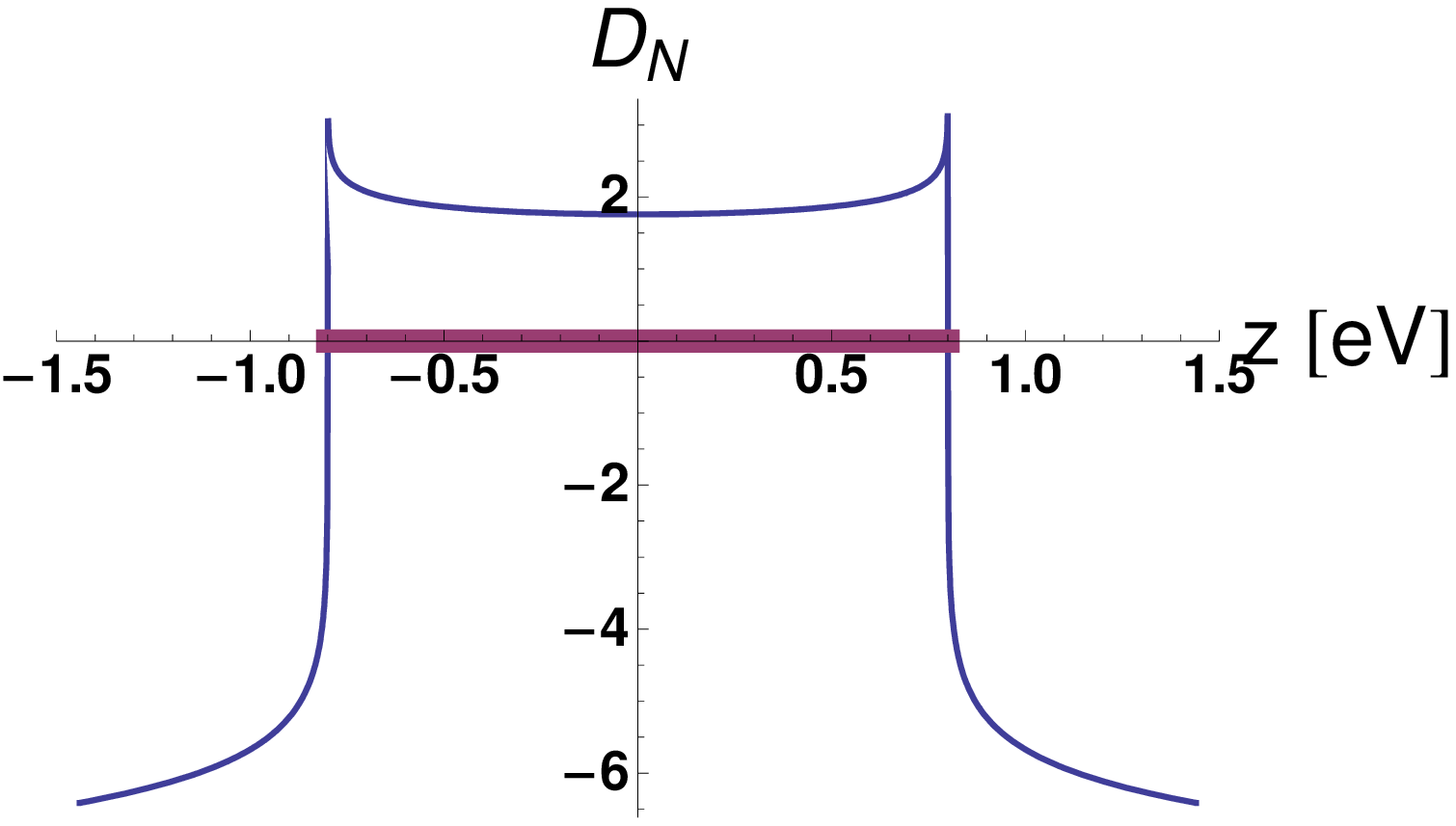,width=4.3cm}

(a) $t_0=0.4$ $v_j=0$

\psfig{file=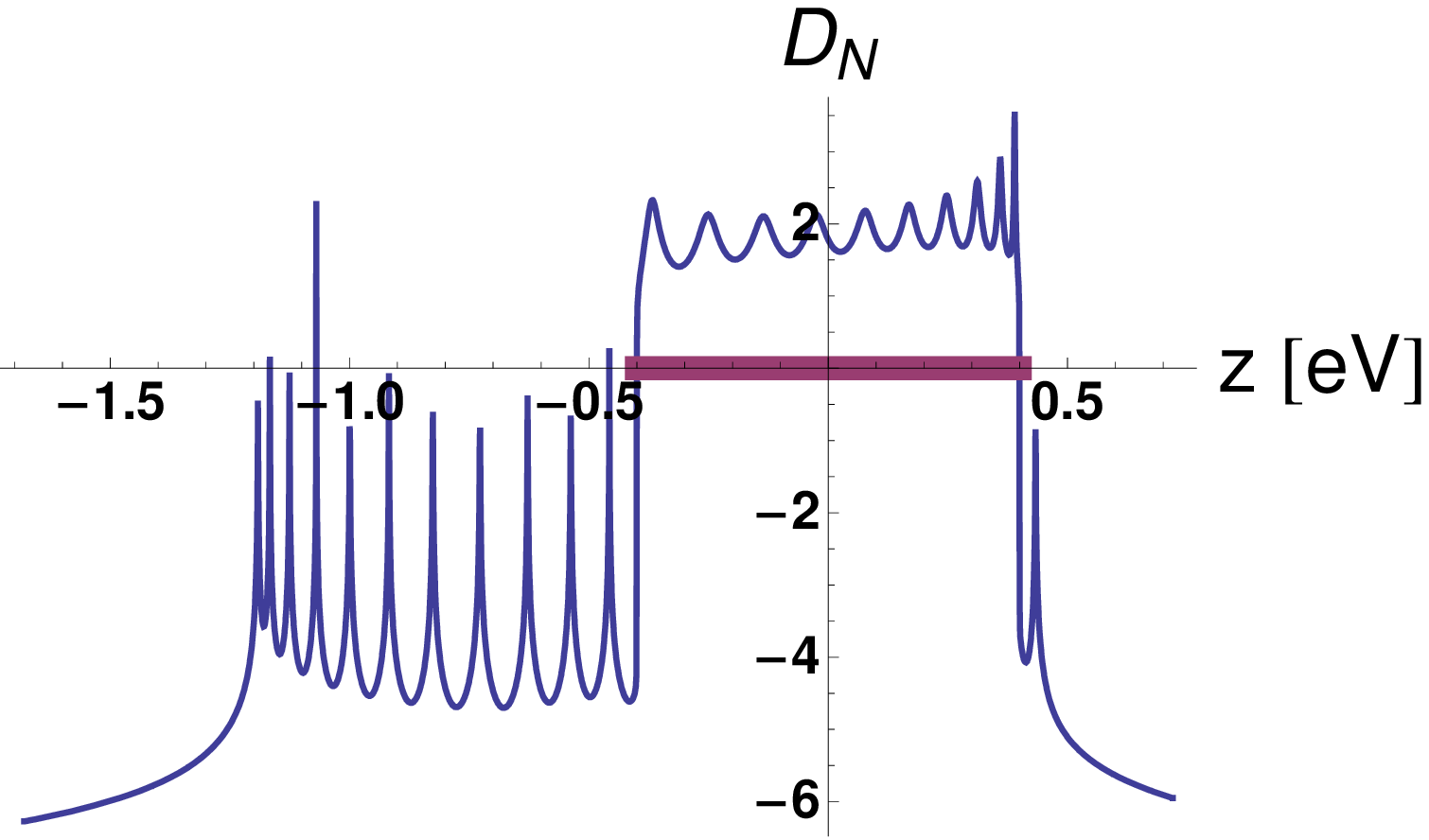,width=4.3cm}

(b) $t_0=0.2$

\psfig{file=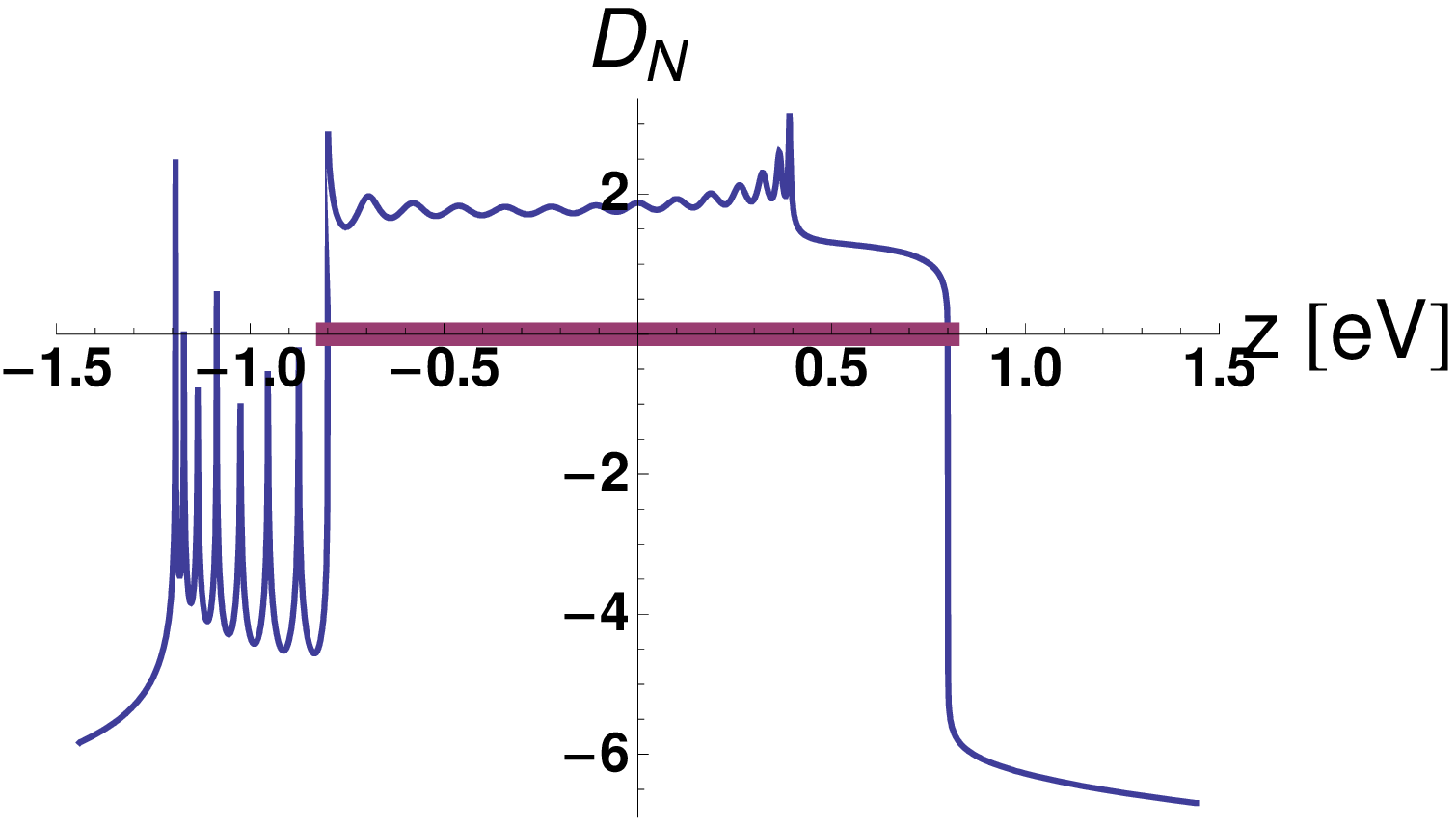,width=4.3cm}

(c) $t_0=0.4$
}
\parbox[]{4.6cm}{
\psfig{file=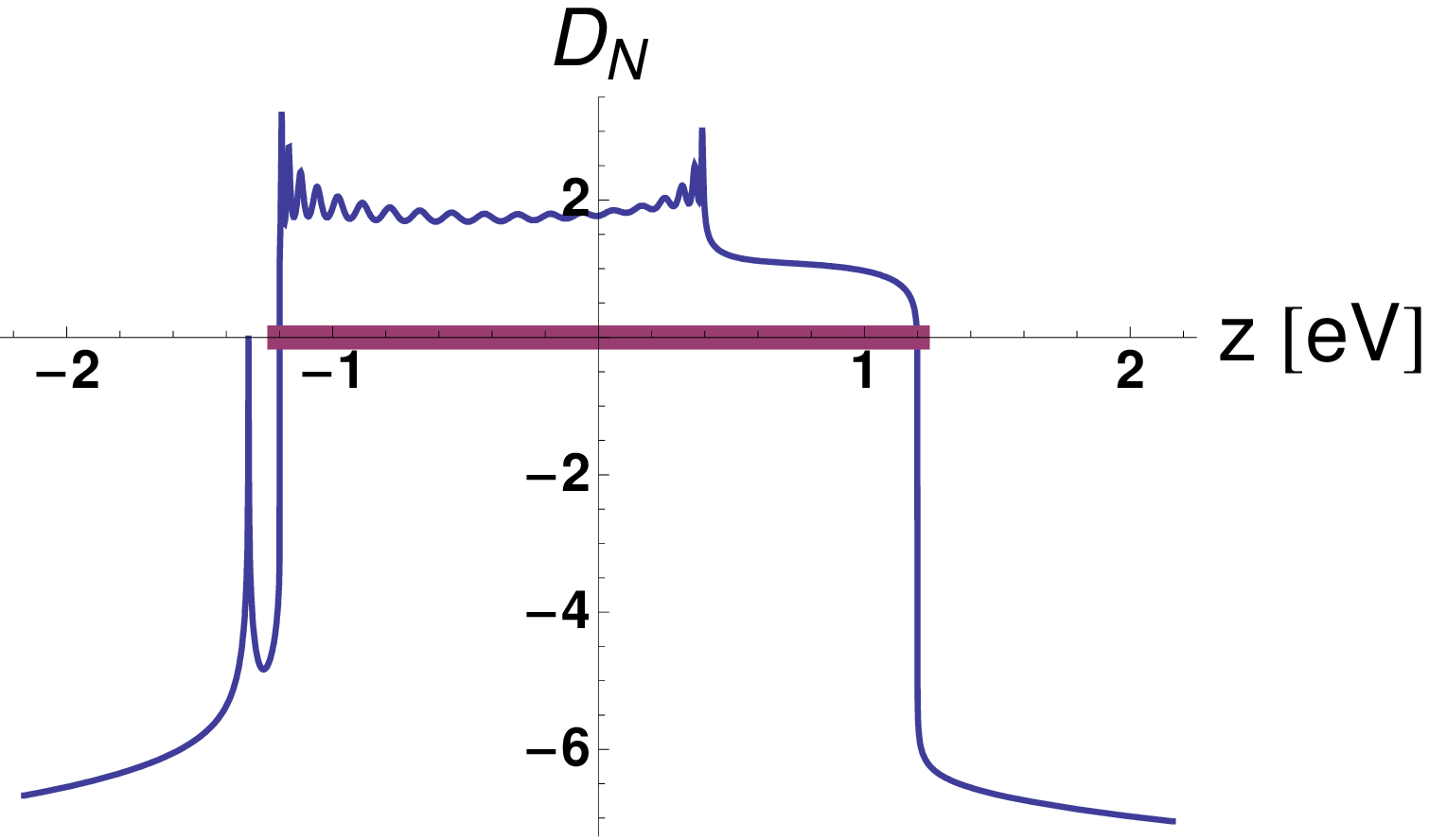,width=4.3cm}

(d) $t_0=0.6$

\psfig{file=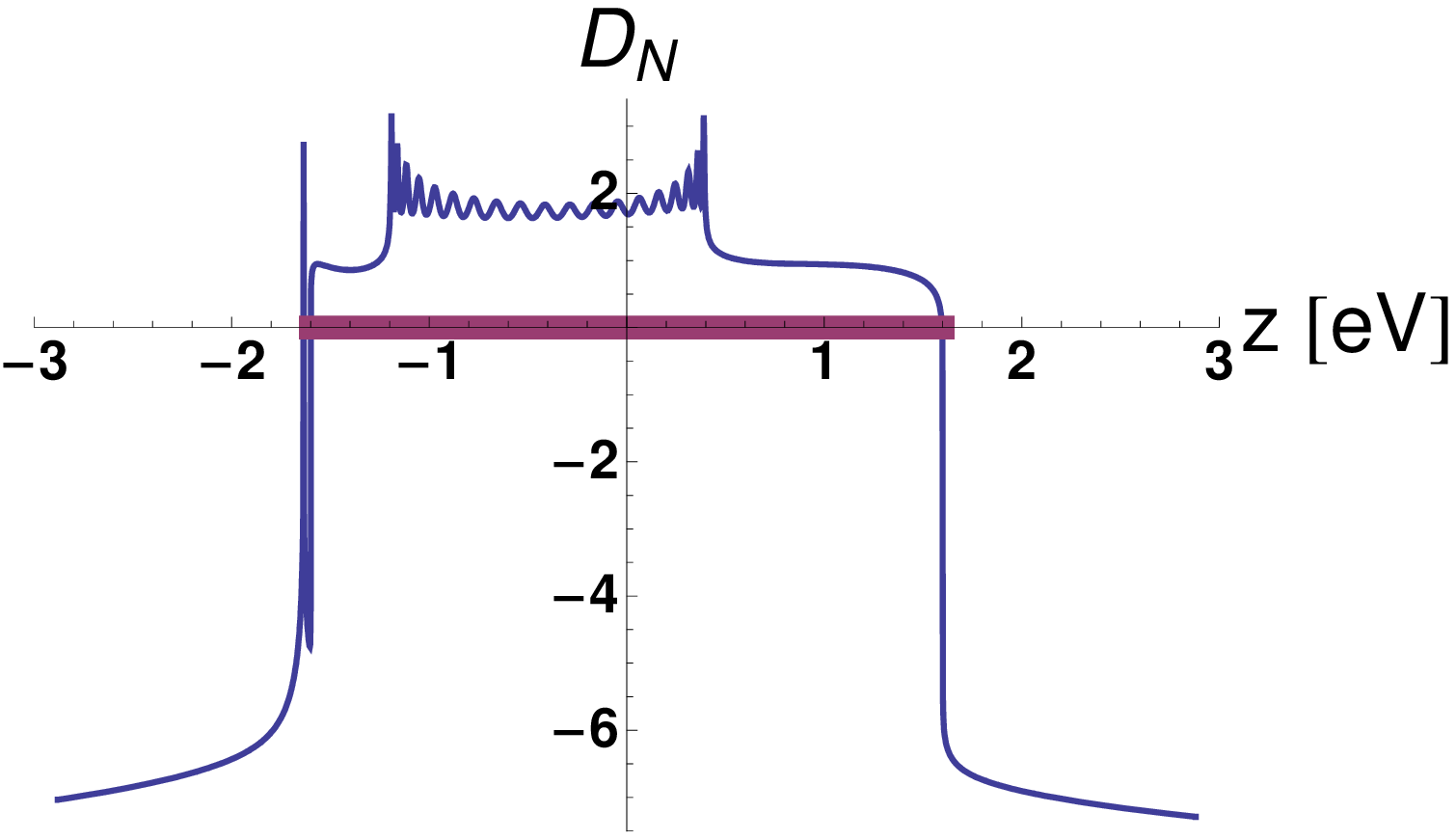,width=4.3cm}

(e) $t_0=0.8$

\psfig{file=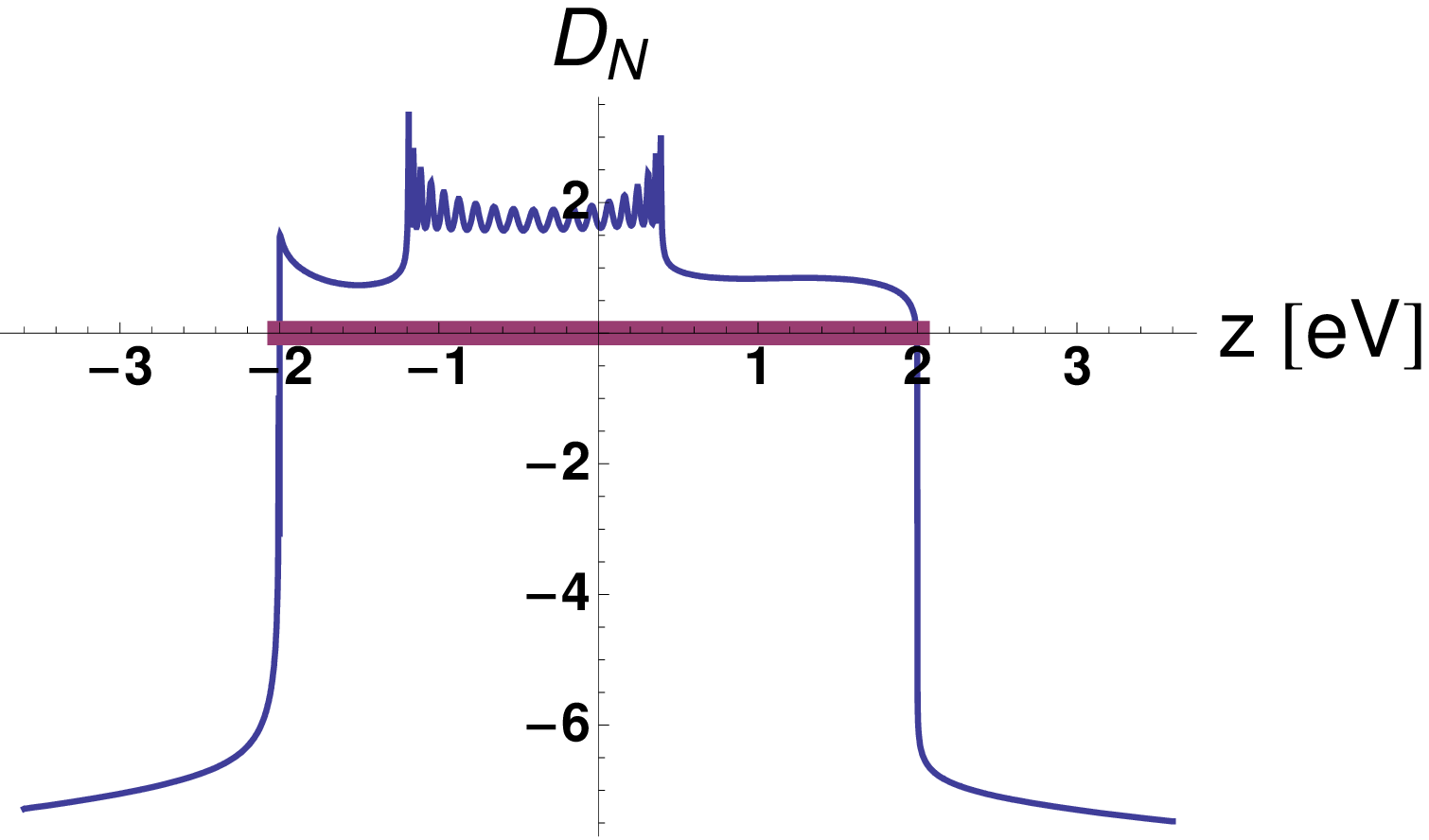,width=4.3cm}

(f) $t_0=1.0$
}
}
\caption{The density of states according to the cases in figure~\ref{g_t0}. The allowed region for transmission is indicated as thick line on the x-axes.\label{dens}}
\end{figure}

The different conductance can be understood from the density of states which is simply given by the diagonal Greenfunction
\be
D_N=\sum\limits_{j=0}^{N+1} {\rm Im} G_{j,j}.
\ee
In figure~\ref{dens} we give the density of states according to figure~\ref{g_t0}. One recognizes the appearance of bound and resonance states below the conductance range for lead hopping parameters less than or equal to the molecular ones, (b) and (c). For larger hopping parameters, (d)-(f), the bound states disappear.
This can be easily understood since electrons are more easily hopping from the leads to the molecules  than transported between the molecules which prevents any standing wave or bound state.

\begin{figure}
\psfig{file=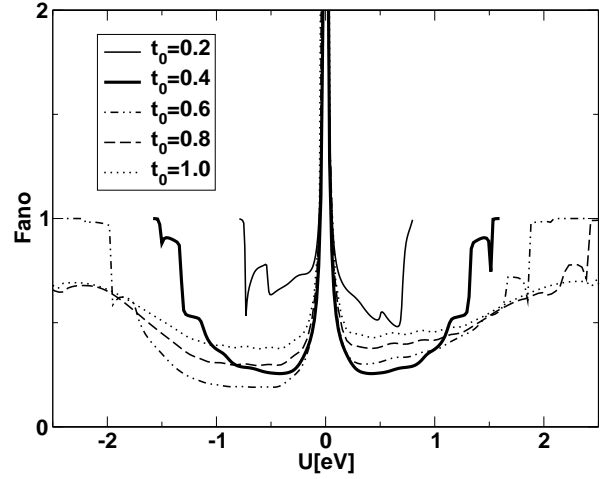,width=9cm}
\caption{The Fanofactor versus applied voltage bias for different hopping parameters as used in figure~\ref{g_t0}.\label{fano}}
\end{figure}

Next we calculate the corresponding current-current fluctuations via the Fano factor (\ref{fano1}). In figure~\ref{fano} we plot the Fano factors corresponding to the situation of figure~\ref{g_t0}.  
One recognizes that the fluctuations are suppressed with increasing lead to molecule hopping. For larger bias voltage, they approach the Schott noise limit, i.e. Poissonian characterized by $F=1$.  It is remarkable that the noise at zero bias becomes super-Poissonian for hopping parameters of the lead smaller than the molecules. For higher hopping parameter these fluctuations are suppressed again.

\begin{figure}
\psfig{file=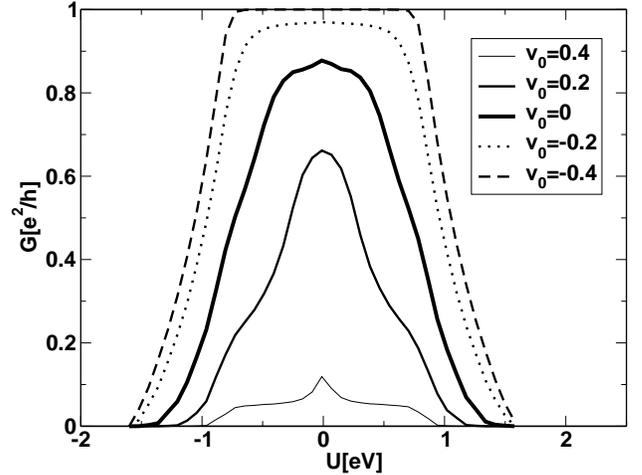,width=9cm}
\caption{The conductance $G$ versus applied voltage bias for different barrier height of the leads. The hopping parameter is $t_0=0.4$ and the thiophene HOMO levels relativ to the leads$v_j=-0.4$. \label{g_v0}}
\end{figure}

As a next question we consider the optimal hopping parameter between Au contacts and the molecules but vary the barrier of the leads. This is motivated by the barrier that occurs at charge injection at metal-organic interfaces described by the image potential \cite{SM99}. In figure~\ref{g_v0} the conductance is plotted for different forms of the lead barrier.

\begin{figure}
\psfig{file=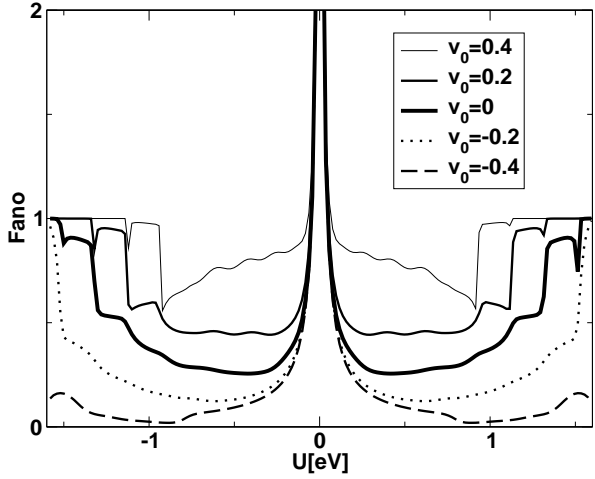,width=9cm}
\caption{The Fanofactor versus applied voltage bias for different barrier heights of the leads as used in figure~\ref{g_v0}.\label{fano_v0}}
\end{figure}

With increasing barrier height the conductance is lowered as one would expect. The corresponding fluctuations in figure~\ref{fano_v0} show that the fluctuations are suppressed with decreasing barrier. This means that the quality of the transport can be improved if the lead to molecule barrier is fabricated as low as possible.

\begin{figure}
\parbox[]{9.5cm}{
\parbox[]{4.6cm}{
\psfig{file=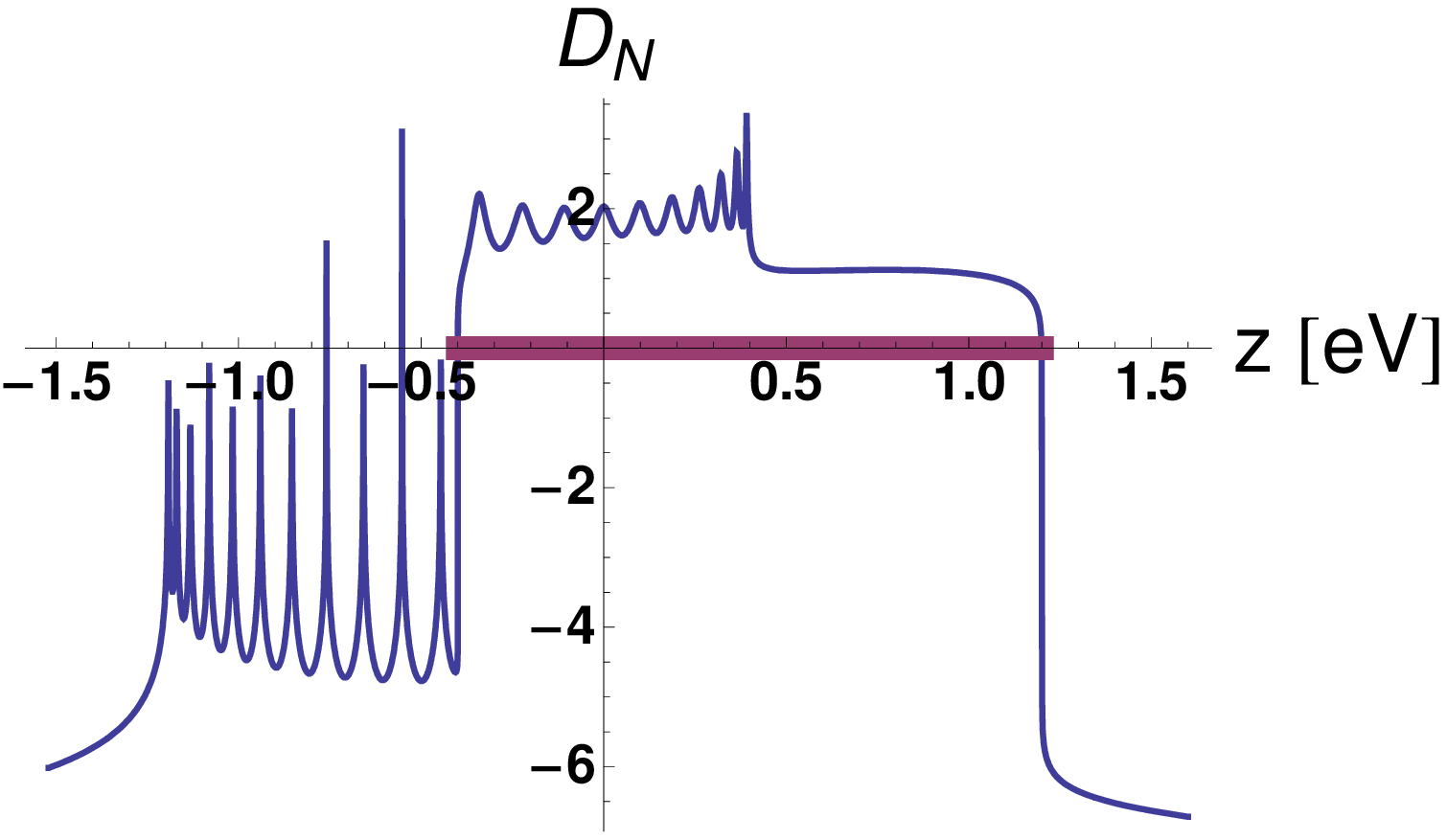,width=4.3cm}

(a) $v_0=0.4$

\psfig{file=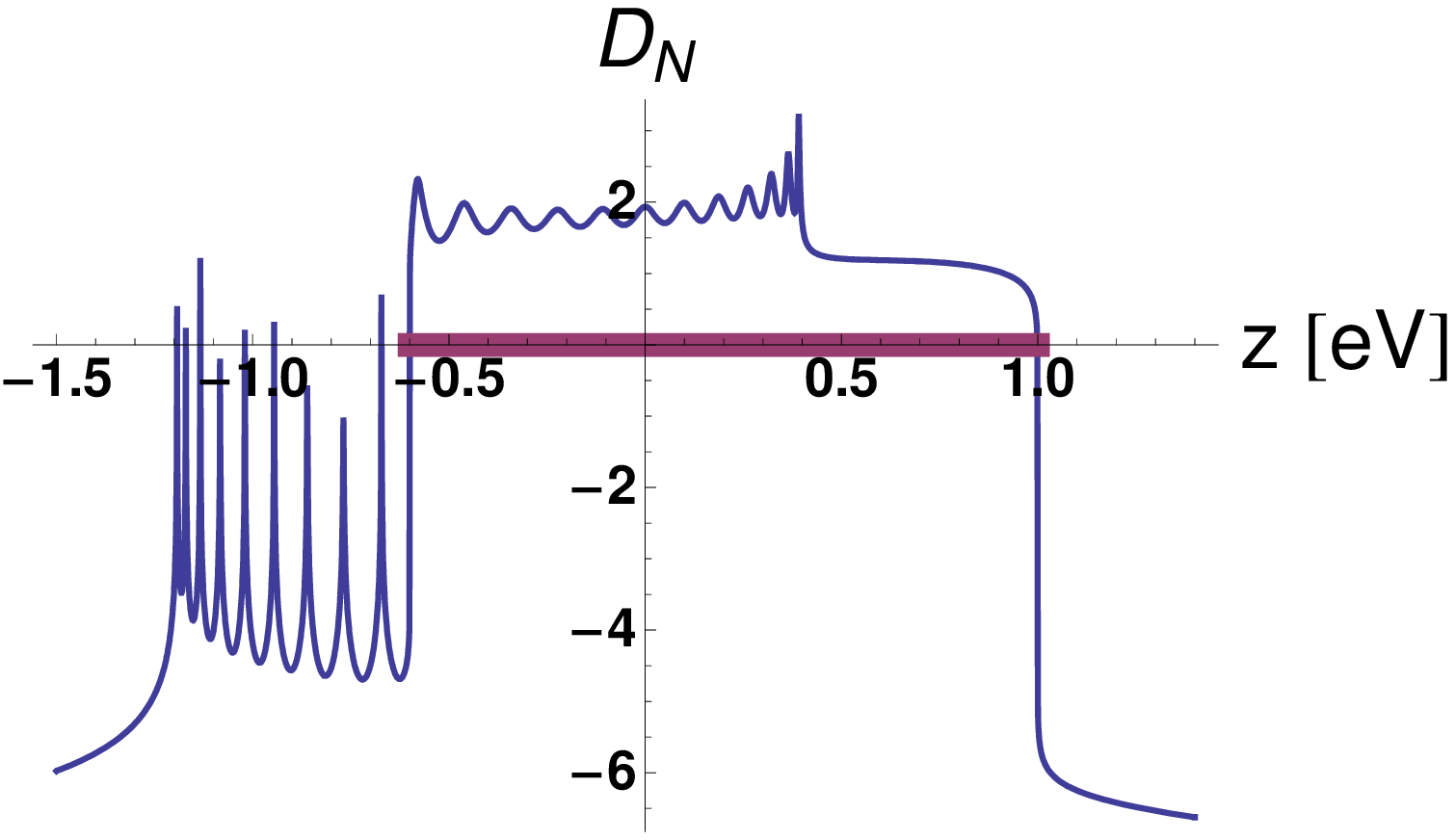,width=4.3cm}

(b) $v_0=0.2$

\psfig{file=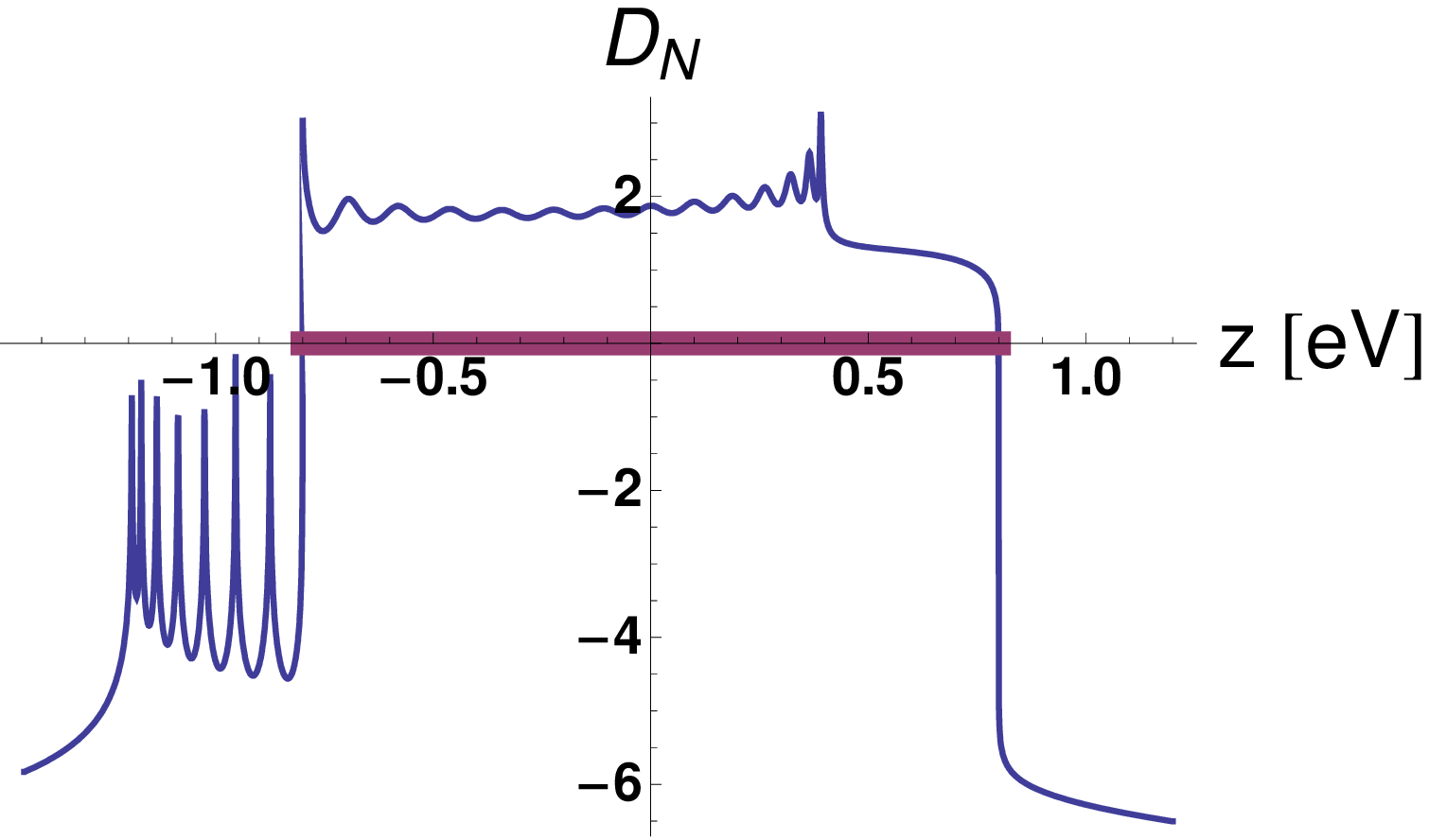,width=4.3cm}

(c) $v_0=0$
}
\parbox[]{4.6cm}{
\psfig{file=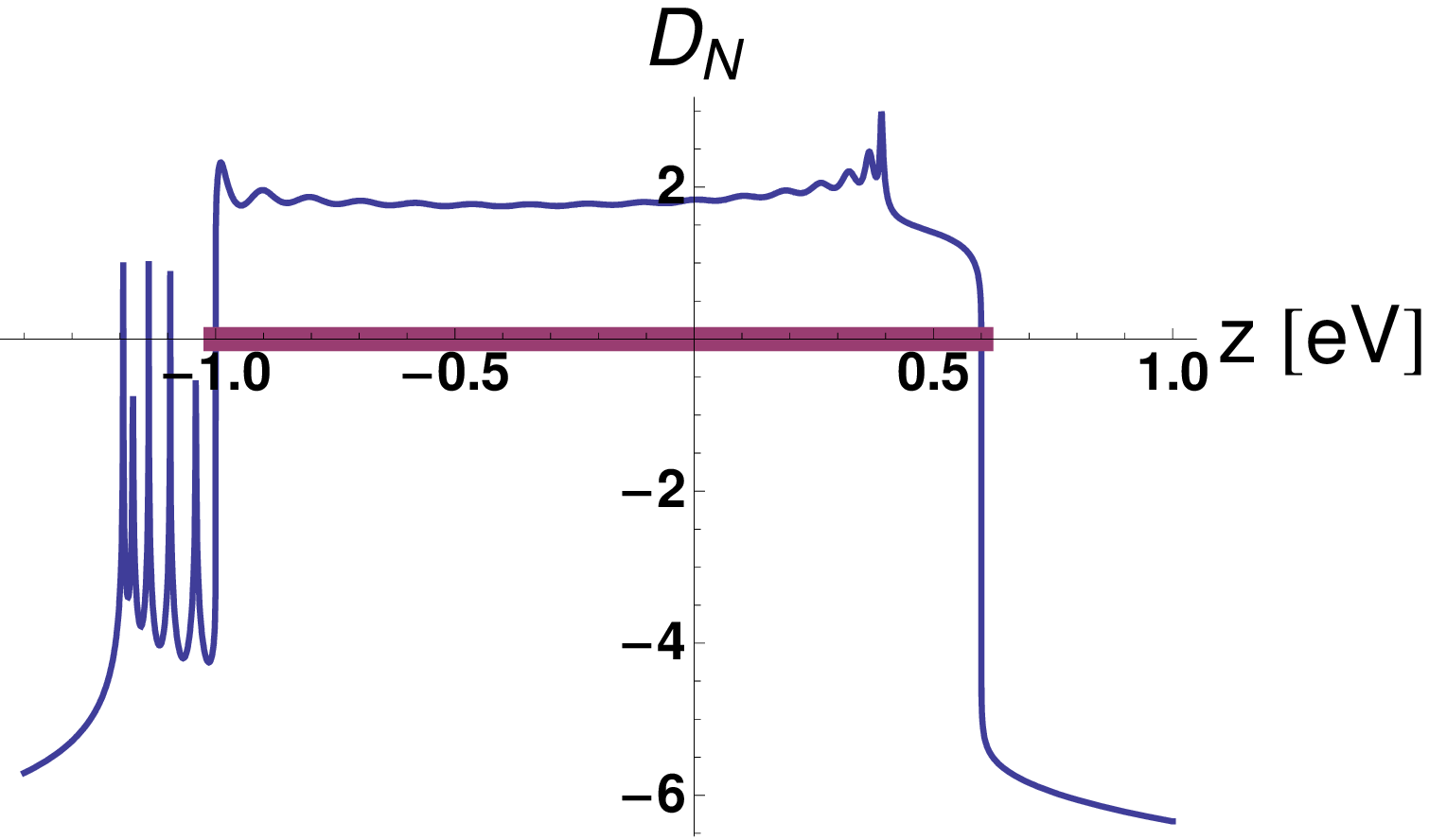,width=4.3cm}

(d) $v_0=-0.2$

\psfig{file=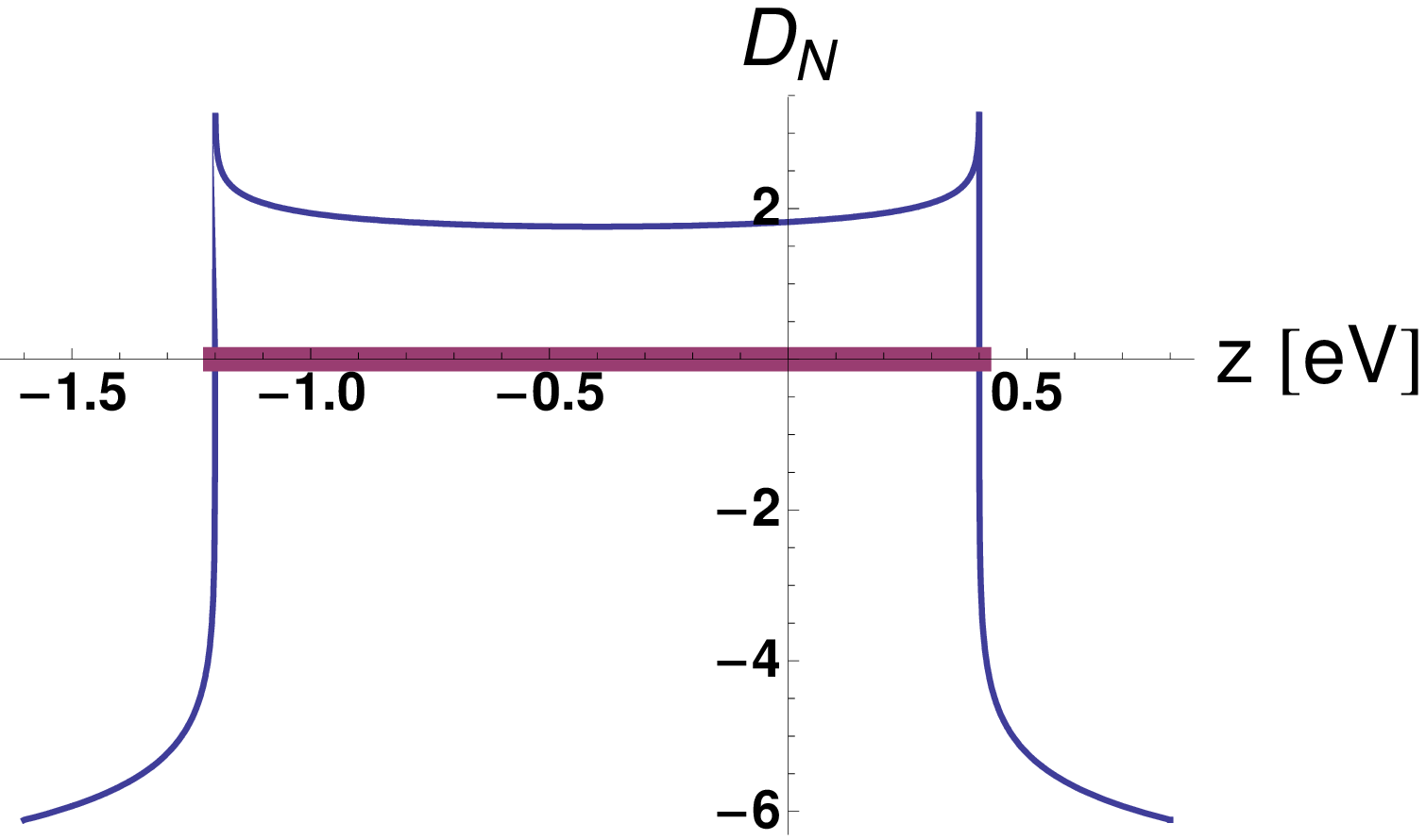,width=4.3cm}

(e) $v_0=-0.4$
}
}
\caption{The density of states according to the figures in figure~\ref{g_v0}. The allowed region for transmission is indicated as thick line on the x-axes.\label{dens_v0}}
\end{figure}

The corresponding density of states in figure~\ref{dens_v0} show that the highest lead barrier (a) yields the largest number of bound states below the conductance range. When lowering the barrier the bound state disappears and the density of states for the almost homogeneous case (e) appears. The remaining difference to the homogeneous case comes from the level mismatch of the HOMO and the leads.

\subsection{Transport coefficients}

With the same tools as used so far it is very easy to compute all transport coefficients, even bias voltage dependent. We assume a different temperature on the right lead $T_r=T+\Delta T/2$ and on the left lead $T_r=T-\Delta T/2$. Linearizing with respect to the temperature gradient, but not with respect to the voltage, the particle current (\ref{jf}) reads
\be
J=L^{11} V+L^{12} \Delta T
\label{o1}
\ee
with the nonlinear Onsager coefficients
\be
L^{11}&=&{e\over \hbar V} \int dz T_{N+1,0} \left [f_r \left ({z-eV\over T} \right )-f_l\left ({z\over T}\right )\right ]
\nonumber\\
L^{12}&=&-{e\over 2 \hbar} \int dz T_{N+1,0} \left [{z-eV\over T^2}{f_r}'-{z\over T^2}{f_l}'\right ]
\ee
with $f'=-f(1-f)$.
With absent  temperature gradients we have the conductance (\ref{G})
\be
G=L^{11}.
\ee

\begin{figure}
\psfig{file=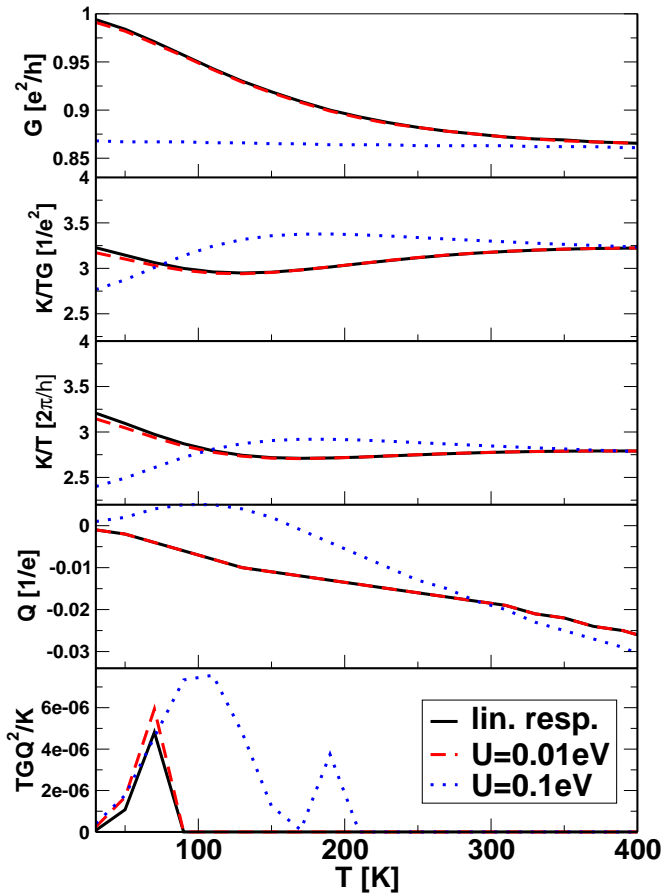,width=9cm}
\caption{The conductance $G$, thermal heat conductance $K$, Wiedemann Franz law, thermopower $Q$ and figure of merit versus temperature for different applied voltage bias near the linear response. \label{all_1}}
\end{figure}

The voltage compensating the current due to the temperature gradient, $V=-Q \Delta T$, such that $J=0$ determines the thermoelectric power (Seebeck coefficient) as
\be
Q={L^{12}\over L^{11}}.
\ee

Analogously we can give the heat current $J_q$ which is just (\ref{jf}) but with an additional energy factor $z$ under the integrand. With small temperature  gradients the heat current takes the form 
\be
J_q=L^{21} V+L^{22} \Delta T
\label{o2}
\ee
with the nonlinear Onsager coefficients
\be
L^{21}&=&{e\over \hbar V} \int dz z T_{N+1,0} \left [f_r \left ({z-eV\over T} \right )-f_l\left ({z\over T}\right )\right ]
\nonumber\\
L^{22}&=&-{e\over 2 \hbar} \int dz z T_{N+1,0} \left [{z-eV\over T^2}{f_r}'-{z\over T^2}{f_l}'\right ].
\ee

The thermoelectric conductance is measured if we keep the particle current zero, $J=0$, which results from (\ref{o1}) and (\ref{o2}) into
\be
J_q=K \Delta T
\ee
with the thermoelectric conductance
\be
K=L^{22}-{L^{21} L^{12}\over L^{11}}.
\ee
The ratio between the thermoelectric conductance and the conductance is called Wiedemann-Franz law and should be linearly proportional to the temperature
\be
{K\over G}={L^{22}\over L^{11}}-{L^{12}L^{21}\over (L^{11})^2}.
\ee

As a measure for the effectiveness of the thermoelectric devices the dimensionless figure of merit or $ZT$ factor is often presented \cite{HD93} as
\be
ZT={T Q^2 G\over K}={T(L^{12})^2\over L^{11}L^{22}-L^{21} L^{12}} 
\ee
which has become an important quantity for thin film thermoelectric devices \cite{V00,VSCQ01}.

\begin{figure}
\psfig{file=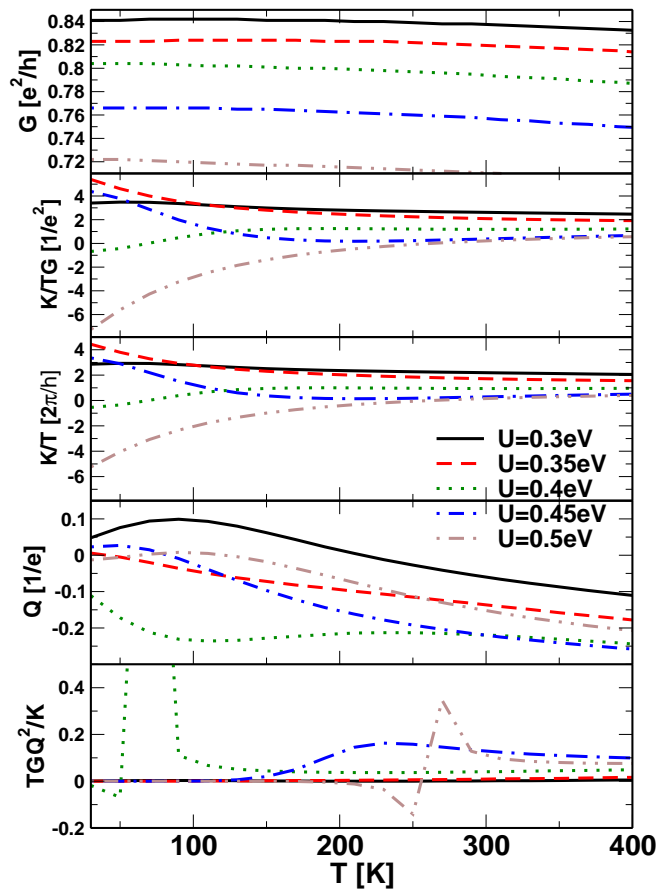,width=9cm}
\caption{The conductance, heat conductance, Wiedemann Franz law, thermopower and figure of merit versus temperature for different applied voltage bias far beyond linear response. \label{all_2}}
\end{figure}

In figures~\ref{all_1} and \ref{all_2} we present the results for the transport coefficients dependent on temperature and the external bias. In figure~\ref{all_1} we plot the results for low external bias to show at which voltage one sees deviations from the linear response result. With increasing applied voltage the conductance is lowered since the overlap between left and right conductance channel shrinks. This was seen already from figures~\ref{g_t0} and \ref{g_v0}.
The thermal conductance shows a nonlinear temperature behavior developing a maximum at certain temperatures. This shows that the thermal conduction is much more sensitive to the interplay between external voltage and HOMO levels of the molecules. The thermopower reveals a similar sensitivity. The Wiedemann Franz law is seen to be not fullfilled strictly. Instead we observe a deviation from the linear temperature behaviour of up to 10\% for lower temperatures. The figure of merit for low applied voltages is negligible but plotted for completeness.

The situation is changing if we apply higher voltages as shown in figure~\ref{all_2}. The observed maximum in the thermal conductance becomes more pronounced and for low temperatures the thermal conductance even changes sign if we approach an external voltage comparable to the HOMO levels. This has dramatic consequences on the figure of merit. At the temperature where the thermal conductance is changing sign we observe a resonance structure enhancing the figure of merit dramatically. For applied voltages equal to the HOMO levels we have the largest resonance but at temperatures around 80K. The resonance is shifted towards higher temperatures for voltages slightly above and  below,  but becomes broader, as well. Interestingly it can be observed that the thermopower is changing sign for certain temperatures and becomes completely negative for voltages equal to the HOMO levels. This shows that for such voltages the thermal current is reversing sign which can be seen analogously to the refrigerator effect described in [\onlinecite{RSKHS07}] by an energy-selective transmission of electrons through a spatially asymmetric resonant structure subject to ac driving.

%\section{Domain switching}

\section{Summary}

Organic molecules (thiophene) ordered paralel between Au contacts are considered under external voltage bias.
The charge transport due to hopping is investigated and the particle and thermal transport properties calculated with the help of the numerically very fast surface Greenfunction technique. A tight binding Hamiltonian is used as underlying model where the hopping parameter are extracted from density functional calculations.

The conductance as well as the current fluctuations can be shaped by changing the lead barrier height. This is due to the bound and resonance states which can be created with high enough barriers. The nonlinear conductance due to high voltage bias is discussed as well in dependence on different hopping parameter which depend strongly on the stacking distance of the molecules. 

We report a staggering of the conductance dependent on the number of molecules squeezed inbetween the two contacts. This finite-number-of-molecule effect we traced down to a coherence phenomenon controlled by the hopping parameter.

The thermal conductance and the thermopower shows a nonlinear behaviour with higher applied voltages even changing sign at special temperatures. This results into a resonance structure in the figure of merit and values near unity. The latter effect was found to be dependent on the applied voltage most pronounced if the voltages is of the order of the energy levels of the HOMO relative to the leads. Besides application of these organic molecules in field effect transitor like structures, these materials might be of interest also for thermoelectric elements.

\acknowledgements

This work was supported by 
the German PPP project of DAAD, by 
DFG 
Priority Program 1157 via GE1202/06 and the BMBF and by European 
ESF program NES as well as
Czech research plans MSM 0021620834 and 
No. AVOZ10100521, by grants GA\v{C}R 202/07/0597 and 202/06/0040 
and GAAV 100100712 and IAA1010404. The financial support by the Brazilian Ministry of Science of Technology is acknowledged.

\appendix
\section{Matrix inversion with surface Greenfunctions}

The method of surface Greenfunctions provides an extremely fast numerical inversion method to solve the Greenfunction equation
\be
G (z-H)=1
\label{green}
\ee
which means the inversion of the matrix $z-H$. Here the Greenfunction are the Fourier transformed causal ones from the time ordering.

We consider the tight-binding Hamiltonian (\ref{ham})
which has the matrix structure
\be
H&=&\left (
\begin{array}{cccc|ccc}
...&&&&&&
\cr
0&t_{j-1}&v_{j-1}&t_j&0&0&0
\cr
0&0&t_{j}&v_{j}&t_{j+1}&0&0
\cr
\hline
0&0&0&t_{j+1}&v_{j+1}&t_{j+2}&0
\cr
0&0&0&0&t_{j+2}&v_{j+2}&t_{j+3}
\cr
&&&&&&...
\end{array}
\right )
\nonumber\\
&&\nonumber\\
&\equiv&\left ( \begin{array}{c|c}
H^{ll} &H^{lr}
\cr
&
\cr
\hline
\cr
H^{rl} &H^{rr}
\end{array}
\right )
\label{matrix}
\ee
where we have cut the matrix at the side $j$.
The cut parts read separately as
\be
H^{rr}&=&\sum\limits_{i=j+1}^{\infty} \biggl ( \ket{i}v_i\bra{i}+\ket{i\!+\!1}t_{i+1}\ket{i}+\ket{i\!-\!1}t_{i}\ket{i\!+\!1} \biggr )
\nonumber\\
H^{ll}&=&\sum\limits_{i=-\infty}^{j} \biggl ( \ket{i}v_i\bra{i}+\ket{i\!+\!1}t_{i+1}\ket{i}+\ket{i\!-\!1}t_{i}\ket{i\!+\!1} \biggr )
\nonumber\\
H^{lr}&=&\ket{j}t_{j+1}\bra{j+1}
\nonumber\\
H^{rl}&=&\ket{j+1}t_{j+1}\bra{j}.
\label{hh}
\ee

Now we employ the general inversion formulas of matrices composed to 2x2 operators, $G B=1$. The upper right equation reads
\be
G_{11} B_{12}+G_{12} B_{22}=0
\ee
from which we obtain 
\be
G_{12}=-G_{11} B_{12} B_{22}^{-1}.
\label{g12a}
\ee 
This is used in the upper left equation of  $G B=1$
\be
G_{11} B_{11}+G_{12} B_{21}=1
\ee
to find finally
\be
G_{11}=\left ({B_{11} -B_{12}B_{22}^{-1}B_{21}}\right )^{-1}
\label{g}
\ee 
and from (\ref{g12a}) the element $G_{12}$ follows. 
The other two elements of $G$ are given by interchanging $1\leftrightarrow 2$.

These formulas allows to write the Greenfunctions (\ref{green}) according to the cutting
(\ref{matrix}) as
\be
%\begin{array}{l @{\hspace{1ex}}l}
G_{11}\!\!&=&\!\!\left ({z\!-\!H^{ll}\!-\!H^{lr} G^{rr} H^{rl}}\right )^{-1}, \quad G_{12}=G_{11} H^{lr} G^{rr},
%\cr
%&
%\cr
\nonumber\\
G_{21}\!\!&=&\!\!G_{22} H^{rl} G^{ll}, \, G_{22}=\left ({z\!-\!H^{rr}\!-\!H^{rl} G^{ll} H^{lr}}\right )^{-1}
%\end{array}
.
\ee
We consider now the specific energy parts in the diagonal Greenfunctions
with the help of (\ref{hh})
\be
H^{lr} G^{rr} H^{rl}&=&\ket{j}t_{j+1}\bra{j+1} G^{rr} \ket{j+1} t_{j+1} \bra{j}
\nonumber\\
&\equiv& t_{j+1}^2 S_{j+1}^r \ket{j}\bra{j}
\ee
which defines the right-side surface Greenfunction $S^r$. Analogously we obtain the left-side Greenfunction
\be
H^{rl} G^{ll} H^{lr}&=&\ket{j+1}t_{j+1}\bra{j} G^{ll} \ket{j} t_{j+1} \bra{j+1}
\nonumber\\
&\equiv& t_{j+1}^2 S_{j}^l \ket{j+1}\bra{j+1}.
\ee

These surface Greenfunction obey simple recursion relations. To see this we cut the left Hamiltonian as
\be
H^{ll}&=&\left (
\begin{array}{cc|c}
...&&0
\cr
t_{j-1}&v_{j-1}&t_j
\cr
\hline
0&t_{j}&v_{j}
\end{array}
\right )
\nonumber\\
&&\nonumber\\
&\equiv&\left ( \begin{array}{c|c}
H^{ll}[j-1] &
\left (\begin{array}{c}
\stackrel{\stackrel{.}{\stackrel{.}{.}}}{0}\cr t_j
\end{array}
\right )
\cr
\hline
\cr
\left (\begin{array}{cc}
...0 & t_j
\end{array}
\right )
&v_j
\end{array}
\right )
\label{matrix1}
\ee
where $H^{ll}[j-1]$ denotes the left-side Hamiltonian when (\ref{matrix}) is cutted at $j-1$.
Due to  the equation $G^{ll} (z-H^{ll})=1$ this partition (\ref{matrix1}) enforces the structure of the Greenfunction
\be
G^{ll}=\left (
\begin{array}{cc}
G^{ll}[j-1] &..
\cr
..&G_{jj}^{ll}
\end{array}
\right ).
\label{matrix2}
\ee
The left upper part is the Greenfunction $G^{ll}$ with the cut at $j-1$.
With the help of the inversion formulas above, the right lower element of (\ref{matrix2})
determines the left-side surface Greenfunction
\be
S_j^l&\equiv& G^{ll}_{jj}={1\over z-v_j-(...0,-t_j) G^{ll}[j-1] \left (\begin{array}{c} \stackrel{\stackrel{.}{\stackrel{.}{.}}}{0}\cr -t_j\end{array} \right )}
\nonumber\\
&=& {1\over z-v_j-t_j^2 S_{j-1}^l}
\label{sl}
\ee
which establishes the recursion formula for the left-side surface Greenfunction. Similiarly one obtains the recursion for the right-side surface Greenfunction
\be
S_{j+1}^r={1\over z-v_{j+1} -t_{j+1}^2 S_{j+2}^r}.
\label{sr}
\ee

If the tight-binding system is in-between two leads with the left material characterized by
$t_j=t_0$ and the crystal levels $v_j=v_0$ for $j\le0$ and the right material characterized by
$t_j=t_{N+1}$ and $v_j=v_{N+1}$ for $j\ge N+1$ one finds from (\ref{sl}) directly the surface Greenfunction of the leads
\be
S_0^l={1\over t_0} \left ({z-v_0\over 2 t_0}-i \sqrt{1-\left ({z-v_0\over 2 t_0}\right )^2} \right )
\label{s0}
\ee
and analogously for $S_{N+1}^r$. Together with (\ref{sl}) this determines all the surface Greenfunctions completely.

With the knowledge of the surface Greenfunction we can now provide the complete Greenfunction (\ref{g}). Employing the familiar relation ${1\over A-B}={1\over A}+{1\over A}B{1\over A}+...$ we can write the matrix elements of 
\be
&&\bra{k} G_{11} \ket{m}=\bra{k} {(1\over (G^{ll})^{-1}-a \ket{j}\bra{j}} \ket{m}
\nonumber\\
&=& \bra{k} G^{ll} + G^{ll} a\ket{j}\bra{j} G^{ll}+G^{ll} a\ket{j}\bra{j} G^{ll} a\ket{j}\bra{j} G^{ll}\nonumber\\&&\qquad +...\ket{m}
\nonumber\\
&=& G_{km}^{ll}+G_{kj}^{ll} a \left (1+S_j^l a+(S_j^l a)^2+...\right ) G_{jm}^{ll}
\nonumber\\
&=& G_{km}^{ll}+{G_{kj}^{ll} aG_{jm}^{ll}\over 1-a S_j^l}
\label{g11}
\ee
where we used the abbreviation $a=S_{j+1}^r t_{j+1}^2$. Analogously we obtain
\be
\bra{k} G_{12} \ket{m}&=&{G_{kj}^{ll} t_{j+1}G_{j+1m}^{rr}\over 1-a S_j^l}
\label{g12}
\\
\bra{k} G_{22} \ket{m}&=&G_{km}^{rr}+{G_{kj+1}^{rr} bG_{j+1m}^{rr}\over 1-b S_{j+1}^r}
\label{g22}
\\
\bra{k} G_{21} \ket{m}&=&{G_{kj+1}^{rr} t_{j+1}G_{jm}^{ll}\over 1-b S_{j+1}^r}
\label{g21}
\ee
with the abbreviation $b=t_{j+1}^2 S_j^l$.

Now we remember the structure of the complete Greenfunction following (\ref{matrix})
\be
\left (
\begin{array}{c|cc}
m&\rightarrow ...j&j+1...
\cr
\hline
\begin{array}{c}
k\cr \downarrow \cr \stackrel{\stackrel{.}{.}}{.} \cr j
\end{array}& G_{11}&G_{12}
\cr
%\hline
\begin{array}{c}
j+1\cr \stackrel{\stackrel{.}{.}}{.}
\end{array}& G_{21}&G_{22}
\end{array}
\right ).
\label{gg}
\ee
Comparing $G_{21}$ of (\ref{g21}) for $k=j+1$ and $m\le j$ in (\ref{gg})
with $G_{11}$ of (\ref{g11}) for $k=j$ and $m\le j$ in (\ref{gg})
we obtain the recursive relation
\be
G_{j+1,m}=t_{j+1} S_{j+1}^r G_{j,m},\qquad m\le j
\label{ga}
\ee
which allows to construct the Greenfunction on one side from its diagonal part. Analogously we obtain the other diagonal part by comparing $G_{22}$ of (\ref{g22}) for $k=j+1$ and $m\ge j+1$ in (\ref{gg})
with $G_{12}$ of (\ref{g11}) for $k=j$ and $m\ge j+1$ in (\ref{gg}) as
\be
G_{j,m}=t_{j+1} S_{j}^l G_{j+1,m},\qquad m\ge j+1.
\label{gb}
\ee
The diagonal form is found from $G_{11}$ of (\ref{g11}) easily to be
\be
\bra{k} G_{11} \ket{k}&=&{S_k^{l}\over 1-t_{k+1}^2 S_{k+1}^r S_{k}^l}
\nonumber\\
&=&{1\over z-v_k-t_{k+1}^2 S_{k+1}^r -t_k^2 S_{k-1}^l}
\label{gc}
\ee
where we used (\ref{sl}) once more.
This completes the recursive construction of the Greenfunction.
\subsection{Summary of the method}
For computational purposes let us collect the important steps and formulas.
\begin{enumerate}
\item
Calculate the surface Greenfunctions of the left and right lead according to (\ref{s0})
\be
S_{N+1}^r&=&{1\over t_{N+1}} \left ({z-v_{N+1}\over 2 t_{n+1}}-i \sqrt{1-\left ({z-v_{N+1}\over 2 t_{N+1}}\right )^2} \right ).
\nonumber\\
S_0^l&=&{1\over t_0} \left ({z-v_0\over 2 t_0}-i \sqrt{1-\left ({z-v_0\over 2 t_0}\right )^2} \right )
\label{sn}
\ee
\item
Determine the surface Greenfunctions due to the recursive relations (\ref{sl}) and (\ref{sr})
\be
S_j^l&=& {1\over z-v_j-t_j^2 S_{j-1}^l}
\nonumber\\
S_{j+1}^r&=&{1\over z-v_{j+1} -t_{j+1}^2 S_{j+2}^r}.
\ee
\item
Compute the required elements of the Greenfunction matrix according to (\ref{ga}), (\ref{gb}) and (\ref{gc})
\be
G_{kk}&=&{1\over z-v_k-t_{k+1}^2 S_{k+1}^r -t_k^2 S_{k-1}^l}
\nonumber\\
G_{j,m}&=&t_{j+1} S_{j}^l G_{j+1,m},\qquad m\ge j+1
\nonumber\\
G_{j+1,m}&=&t_{j+1} S_{j+1}^r G_{j,m},\qquad m\le j.
\ee
\end{enumerate}
                                
\vspace{-3ex}

\bibliography{kmsr,kmsr1,kmsr2,kmsr3,kmsr4,kmsr5,kmsr6,kmsr7,delay2,spin,gdr,refer,sem1,sem2,sem3,micha,genn,solid,shuttling}
\end{document}